\documentclass[superscriptaddress,showpacs,nofootinbib,twocolumn,pra]{revtex4-1}

\usepackage{amsmath,amsthm,amssymb,bm,hyperref,graphicx,color}
\usepackage[toc,page]{appendix}

\newcommand{\be}{\begin{equation}}
\newcommand{\ee}{\end{equation}}

\newcommand{\inti}{\int_{-\infty}^{+\infty}}

\newcommand{\6}{\partial}

\begin{document}

\title{Temperature driven crossover in the Lieb-Liniger model}

\author{Andreas Kl\"umper}
\affiliation{Fachbereich C – Physik, Bergische Universit\"at  Wuppertal,
42097 Wuppertal, Germany}

\author{Ovidiu I.~P\^{a}\c{t}u}
\affiliation{Institute for Space Sciences, Bucharest-M\u{a}gurele, R
077125, Romania}


\begin{abstract}

The large-distance behavior of the density-density correlation function in the
Lieb-Liniger model at finite temperature is investigated by means of the recently
derived  nonlinear integral equations characterizing the correlation lengths.
We present extensive numerical results covering all the physical regimes from
 weak to strong interaction  and all temperatures. We find that the leading
term of the asymptotic expansion becomes oscillatory at a critical temperature
which decreases with the strength of the interaction. As we approach the
Tonks-Girardeau limit the asymptotic behavior becomes more complex  with
a double crossover of the largest and next-largest correlation lengths.
The crossovers exist only for intermediate couplings and vanish
for $\gamma=0$ and $\gamma=\infty$.

\end{abstract}

\maketitle


Correlation functions play a fundamental role in our understanding
of low-dimensional strongly correlated systems. As a result of the
remarkable progress in the field of ultracold gases  atomic correlations
can now be accessed through a variety of techniques \cite{CCGOR,GBL}
(and references therein) highlighting  the necessity of high-quality
experimental phenomenology. In one dimension the class of exactly solvable
models represent a  particular type of strongly correlated systems for
which the powerful techniques associated with Bethe ansatz allows us
to go beyond the mean-field  Bogoliubov approximation. The paradigmatic
example is the Lieb-Liniger (LL)  model \cite{LL}
which is experimentally realizable \cite{tMSK,tKWW1,tP,tLOH,tPRD,tAetal}
and has been the subject of various theoretical investigations for more than
fifty years \cite{KBI}.

In this article we investigate the large distance of the static density-density
correlation function at finite temperature in the LL model. Even though the
model is integrable the complicated form of the wavefunctions means that the
analytical derivation or numerical analysis of the correlators is still an
extremely challenging task. While important progress has been made in recent
years
\cite{BK1,IK1,Ka,KGDS,CB,CC,SGDVK,DSGDDK,KCI,BAKG,WHLYGB,PC,FPCF,KMS1,PK} a
complete characterization of the temperature dependent correlators is still
lacking.   At zero temperature there are two distinct
  phases in the LL model: for chemical potential $\mu<0$ the particle density
  is zero, whereas $\mu>0$ is characterized by a finite density and realizes a
  Tomonaga Luttinger liquid (TLL) with algebraically decaying correlation
  functions that are described by Conformal Field Theory (CFT). At low but
  finite temperature the TLL phase is still present with exponentially
  decaying correlation functions and correlation lengths determined by CFT.
  The phase $\mu<0$ may be denoted `gas phase' as at low temperatures it has a
  non-zero but small density of particles and the ideal gas law holds. The
  main interest of this article lies on the finite density  phase at finite
  especially intermediate and high temperatures, where the correlation
  functions are not described by TLL/CFT and which has not been thoroughly
  investigated until now.  Our analysis is based on the use of the so-called
quantum transfer matrix $T_q$ which describes the evolution of correlators in
spatial direction. The eigenvalue equations \cite{KMS1,PK} enjoy a symmetry
property which is equivalent to saying that $T_q$ is a normal operator,
i.e. $[T_q^+,T_q]=0$ from which follows that all eigenvalues are real or occur
in complex conjugate pairs. We calculated the  leading
  and next-leading eigenvalues  of $T_q$ for various densities and
temperatures and found results which cannot be captured by the
Tomonaga-Luttinger liquid/Conformal Field Theory (TLL/CFT) \cite{CFT} approach
or other approximation or numerical methods. There is a complex crossover
scenario with a counterintuitive change of symmetry and dimensionality of the
involved states.  In the spectral decomposition of the density-density
correlator the leading state is symmetric at low temperatures, but has broken
symmetry at higher (!)  temperatures where it is 2-dimensional with complex
conjugate eigenvalues.

{\it The model - } The Lieb-Liniger model describes one-dimensional bosons
interacting via a $\delta$-function potential. The second-quantized Hamiltonian
is
\be\label{hamLL}
H=\int\,  dx\,  \6_x\Psi^\dagger(x)\6_x\Psi(x)+
c\Psi^\dagger(x)\Psi^\dagger(x)\Psi(x)\Psi(x)\, ,
\ee
where $c>0$ is the coupling constant and we consider $\hbar=2m=k_B=1$,
with $m$ the mass of the particles. In (\ref{hamLL}) $\Psi^\dagger(x)$
and $\Psi(x)$ are  Bose fields satisfying the canonical commutation
relations $ [\Psi(x),\Psi^\dagger(x')]=\delta(x-x')\, ,\ \ \ \
[\Psi(x),\Psi(x')]=[\Psi^\dagger(x),\Psi^\dagger(x')]=0\, .$ The LL
model is exactly solvable \cite{LL,KBI,YY4} and, at finite temperature,
is completely characterized by two parameters: the coupling strength
$\gamma=c/n$  with $n$ the linear density and the temperature $T$.

The density-density correlation function is defined by
\[
\langle \rho(x)\rho(0)\rangle_T=\frac{\textsf{Tr }
[e^{-H/T}\rho(x)\rho(0)]}{\textsf{Tr } e^{-H/T}}\, ,
\]
where $\rho(x)=\Psi^\dagger(x)\Psi(x)$. Due
to translational invariance and invariance under spatial reflection
$\langle \rho(x)\rho(0)\rangle_T=\langle \rho(0)\rho(x)\rangle_T$
which means that it is sufficient to consider $x> 0$.  The density-density
correlator is closely related to the (unnormalized) second order correlation function
$g^2_T(x)=\langle :\rho(x)\rho(0):\rangle_T$  via  $\langle \rho(x)\rho(0)\rangle_T=g^2_T(x)+\delta(x) n$
where  $:\ :$ denotes normal ordering.
The large-distance asymptotic expansion of $\langle \rho(x)\rho(0)\rangle_T$ valid for any value
of the coupling strength and temperature has  been derived  only recently
\cite{KMS1,PK}:
\be\label{AEdd}
\langle \rho(x)\rho(0)\rangle_T= n^2+\sum_i  A_i\,
  e^{-\frac{x}{\xi[u_i]}}\, ,
\ \ \ \ \ x\rightarrow\infty\, ,
\ee
with  the correlation lengths given by
\begin{align}\label{cldd}
\frac{1}{\xi[u_i]}=\frac{1}{2\pi}&\int_{\mathbb{R}}\log
\left(\frac{1+\, e^{-\varepsilon(k)/T}}{1+e^{-u_i(k)/T}}
\right)\, dk\nonumber\\
&-i\sum_{j=1}^r k^+_j+i\sum_{j=1}^r k^-_j\, ,
\end{align}
and $\varepsilon(k)$  the excitation energy satisfying
the Yang-Yang equation
\be\label{dressede}
\varepsilon(k)=k^2-\mu-\frac{T}{2\pi}\int_{\mathbb{R}}
K(k-k')\log
\left(1+e^{-\varepsilon(k')/T}\right)dk'\, .
\ee
In Eq.~(\ref{AEdd}) $u_i(k)$ are a set of functions satisfying the nonlinear
integral equations (NLIEs)
\begin{align}\label{unls}
u_i(k)&=k^2-\mu+iT\sum_{j=1}^r\theta(k-k^+_j)
-iT\sum_{j=1}^r\theta(k-k^-_j)\nonumber\\
&-\frac{T}{2\pi}\int_{\mathbb{R}}K(k-k')
\ln\left(1+e^{-u_i(k')/T}\right)dk'
\, ,
\end{align}
where $\theta(k)=i\ln\left(\frac{ic+k}{ic-k}\right)\, ,
\mbox{lim}_{k\rightarrow\pm\infty}\theta(k)=\pm \pi\, $ and $
K(k-k')=\frac{d}{dk}\theta(k-k')=2c/[(k-k')^2+c^2]\, .$ Eq.~(\ref{unls})
depends on $2r$ parameters, $\{k^+_j\}_{j=1}^r\,  (\{k^-_j\}_{j=1}^r) $
located in the upper (lower) half of the complex plane which are subject
to the  constraint
\be\label{constr}
1+e^{-u_i(k^\pm_j)/T}=0.
\ee
For a given $r$ there are more than one set of $\{k_j^\pm\}_{j=1}^r$
that satisfy the previous constraint and each one defines a distinct
$u$ function. The subscript $i$ labels all these functions for all
$r=1,2,\cdots$.

 It should be noted that the correlation lengths
  $\xi[u_i]$ can be complex, but then appear in conjugate pairs. In this case
  the two appropriate terms in the expansion (\ref{AEdd}) oscillate and may be
  combined into $\mbox{Re }[ A_i\, e^{-x/\xi[u_i]}]$. We should also make an
observation regarding the  prefactors $A_i$ appearing
in (\ref{AEdd}). While analytical expressions were derived in \cite{KMS1} they
are too cumbersome to allow for an efficient numerical investigation.  For
this reason in the following we are going to consider these prefactors as
unknowns. Eq.~(\ref{AEdd}) can be understood as the generalization for all
non-zero temperatures of the TLL/CFT asymptotic expansion \cite{CFT} (valid
for low-T and $x\gg n^{-1}$)
\begin{align}\label{inte1}
&\langle \rho(x)\rho(0)\rangle_T-n^2=-
\frac{(T\mathcal{Z}/v_F)^2}{2\sinh^2(\pi T x/v_F)}\nonumber\\
&\ \ \ \  \ \  +\sum_{l\in\mathbb{Z}^*}\ A_l\ e^{2i xl k_F }
\left(\frac{\pi T/v_F}{\sinh(\pi T x/v_F)}
\right)^{2l^2\mathcal{Z}^2}\, .
\end{align}
In (\ref{inte1}), $v_F$ and $k_F$ are the Fermi velocity and wavevector, $A_l$
are unknown prefactors and $\mathcal{Z}=Z(q)$ with the dressed charge $Z(k)$
satisfying the integral equation
$ Z(k)-\frac{1}{2\pi}\int_{- q}^{ q } K(k-k')Z(k')\, dk'=1.$ In \cite{KMS1,PK}
it was shown that in the low temperature limit the correlation lengths obtained using
 (\ref{cldd}) reproduce the TLL/CFT predictions (\ref{inte1}).
We should stress that even though the TLL/CFT expansion (\ref{inte1}) is valid for
$0<\gamma<\infty$  it does not show any crossovers in $T$  as the correlation
lengths are reciprocal to the temperature in the conformal regime.

{\it Tonks-Girardeau limit -} We can perform an additional check of Eq.~(\ref{cldd})
in the limit of impenetrable particles. This will also allow us to understand the
distribution in the complex plane of the discrete parameters that characterize the
leading correlation lengths. In the Tonks-Girardeau limit the second
order correlation function can be calculated analytically \cite{KBI} with the result:
\be\label{i2}
\langle:\rho(x)\rho(0):\rangle_T=n^2-\frac{1}{4\pi^2}\left(\inti e^{i kx}\vartheta(k)
\, dk\right)^2\, ,\ \ \
\ee
where $\vartheta(k)=1/(1+e^{(k^2-\mu)/T})$ is the Fermi function and $n=\frac{1}
{2\pi}\inti \vartheta(k)\, dk$.  As we will show below, even though
(\ref{i2}) is valid for all temperatures, in the limit of impenetrable particles
no crossover is present.

The large distance behavior of (\ref{i2}) (note
that  the density-density correlation function has the same large-distance
asymptotic behavior  as $\langle :\rho(x)\rho(0):\rangle_T$
because they differ only by a delta function
which is zero for large $x$) can be derived deforming
the contour of integration in the upper-half plane (for $x>0$) with the leading terms
being the residues closest to the real axis (see Appendix \ref{aTG}). The poles of the integrand are given by
the solutions of the equation $k^2-\mu=i\pi (2s+1) T\, ,\  s=0,\pm 1,\cdots$. Explicitly,
the solutions  closest to the real axis ($s=0$) are
\begin{align}\label{k}
k_r^\pm&=[(\alpha+\mu)^{1/2}\pm i \left(\alpha-\mu\right)^{1/2}]/\sqrt{2}\, ,\nonumber\\
k_l^\pm&=[-(\alpha+\mu)^{1/2}\pm i \left(\alpha-\mu\right)^{1/2}]/\sqrt{2}\, ,
\end{align}
with $\alpha=\sqrt{\mu^2+\pi^2T^2}$ implying the asymptotic expansion
\begin{align}\label{i3}
\langle \rho(x)\rho(0)\rangle_T&= n^2+\frac{T^2}{2k_l^+k_r^+}e^{i(k_l^++k_r^+)x}+\frac{T^2}{4(k_l^+)^2}e^{2ik_l^+x}\nonumber\\
&\ \  +\frac{T^2}{4(k_r^+)^2}e^{2ik_r^+x}+o(e^{2i \mbox{Re}(k_l^+) x})\, .
\end{align}

We note that
the leading term
of the expansion is oscillatory at all temperatures and no crossover is present.
We can show that the  correlation lengths appearing
in (\ref{i3}) can also be derived using Eq.~(\ref{cldd}). For this, we need to notice that in the
Tonks-Girardeau limit Eq.~(\ref{unls}) reduces to $u_i(k)=k^2-\mu$ which means
that (\ref{constr}) is equivalent to $(k_j^\pm)^2-\mu=i\pi (2s+1) T\, ,\  s=0,\pm 1,\cdots$
and the expression for the correlation lengths takes the simple form
$1/\xi[u_i]=-i\sum_{j=1}^r k^+_j+i\sum_{j=1}^r k^-_j\, .$  Then, the leading
correlation lengths  $(r=1)$ are:  $1/\xi_0=-ik_l^++ik_l^-=-ik_r^++ik_r^-=-i(k_l^++k_r^+)$
and $1/\xi_{-1}=-ik_l^++ik_r^-=-2ik_l^+\, ; 1/\xi_1=-ik_r^++ik_l^-=-2ik_r^+\,  $
proving our previous assertion. In the low-T limit the expansion (\ref{i3})
reduces to the $l=0,\pm 1$ terms of the TLL/CFT expression (\ref{inte1}). This can
be easily seen using $k_r^+=\sqrt{\mu}+i\frac{\pi T}{2\sqrt{\mu}}+O(T^2)\, ,$
$k_l^+=-\sqrt{\mu}+i\frac{\pi T}{2\sqrt{\mu}}+O(T^2)\, ,$ and the fact that for
the impenetrable gas $\mathcal{Z}=1\, ,k_F=\sqrt{\mu}\, ,v_F=2\sqrt{\mu}\, $.

\begin{figure}[h]
\includegraphics[width=\linewidth]{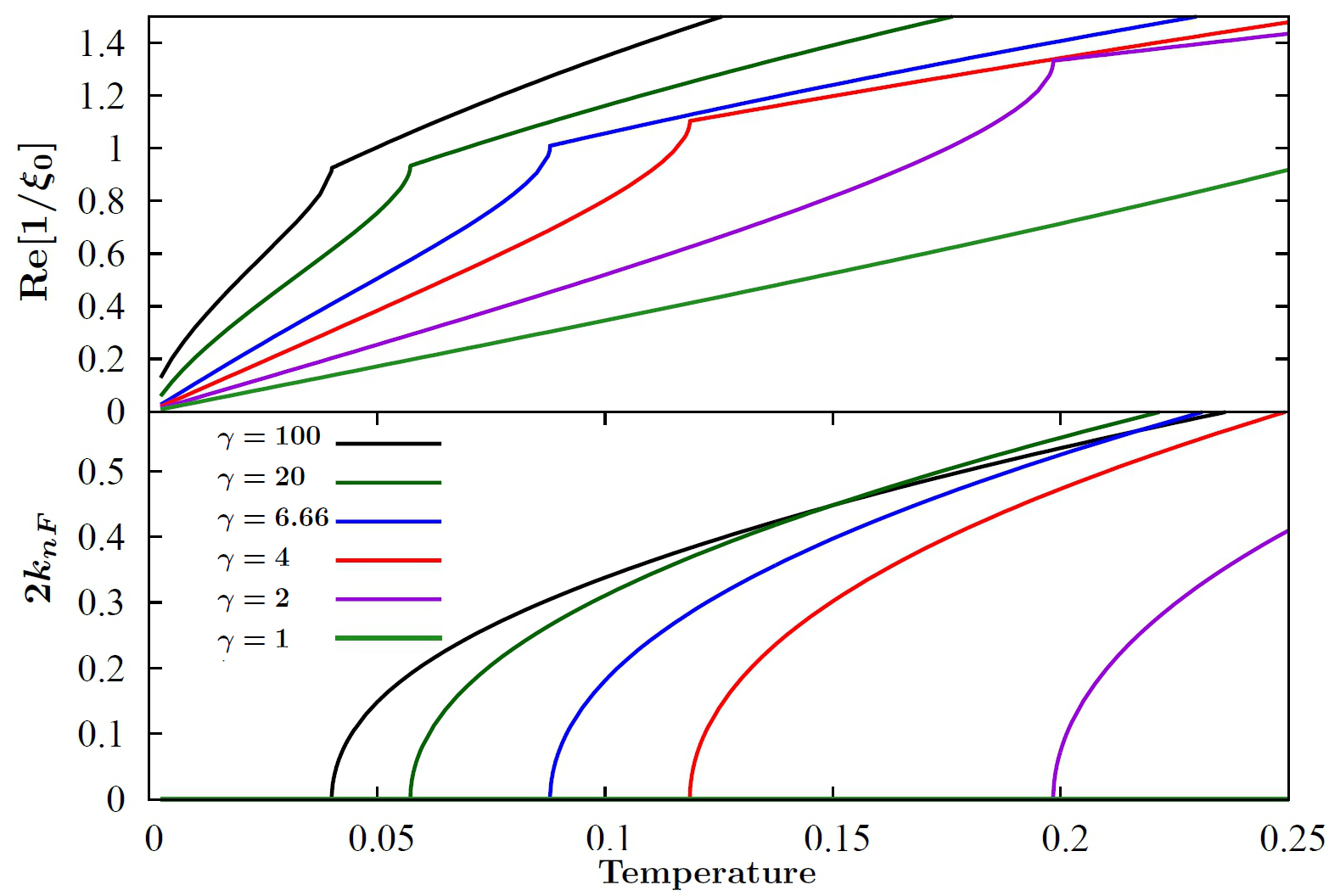}
\caption{(Color online) Upper panel:
$\mbox{Re } [1/\xi_0]$ with $\xi_0$ the leading correlation length as a function of temperature for
various values of $\gamma$ .
Lower panel: $2k_{nF}=\mbox{Im } [1/\xi_0]$  for the same parameters. The critical temperature
$T_o(\gamma)$ is the value for which $2k_{nF}$ becomes nonzero. Note the kinks of   $\mbox{Re } [1/\xi_0]$
at $T_o$.
(All quantities in units of $1/l_0$ and $T_0$ \cite{units}.)}
\label{lead}
\end{figure}

{\it Numerical results - } From the numerical point of view the relevant NLIEs   and  the subsidiary
equations for the discrete parameters  are solved  using an iterative algorithm
which is  quickly convergent. The  efficiency of  the algorithm is enhanced by
the calculation of  convolution  type integrals in   "momentum space" where they
are reduced to products of Fourier transforms  of the involved functions. For all
the functions investigated $\mbox{Re }(1/\xi[u_i])\geq 0$ (a nonzero imaginary part
of the correlation length means that the corresponding term in the expansion is
oscillatory).

The leading correlation length, denoted by $\xi_0$, is
obtained considering  $r=1$  in Eq.~(\ref{unls}) and $k^\pm_1$ the pair of parameters
which  are closest to the real axis with  $k^+_1$  located in the first quadrant of
the complex plane ($\mbox{Re } k^+_1\geq 0\, ,\  \mbox{Im } k^+_1\geq 0$) and
$k^-_1$ located in the  fourth quadrant ($\mbox{Re } k^-_1\geq 0\, ,\  \mbox{Im }
k^-_1\leq 0$.) Similar to the impenetrable case there is another correlation length
with the same magnitude which can be obtained considering $k^+_1$  located in the second quadrant
of the complex plane ($\mbox{Re } k^+_1\leq 0\, ,\  \mbox{Im } k^+_1\geq 0$) and
$k^-_1$ located in the  third quadrant ($\mbox{Re } k^-_1\leq 0\, ,\  \mbox{Im } k^-_1\leq 0$.)
In Fig. \ref{lead} we plot $1/\xi_0$ as a function of temperature
for several values of $\gamma$. At low-T $k^+_1$ and $k^-_1$ are complex conjugate
$(k^+_1=\overline{ k^-_1})$ which also means that $\xi_0$ is real. In this regime
$1/\xi_0$ reproduces the exponential decay predicted by the first term  on the r.h.s
of the TLL/CFT expansion (\ref{inte1}). However, there is
a critical temperature, which depends on $\gamma$, at which the  discrete parameters are no longer
complex conjugates and  $\mbox{Im } [1/\xi_0]$ is no longer zero. Above this temperature,
which we will denote $T_o(\gamma)$, the leading term of the asymptotic expansion becomes
oscillatory with a  wavevector $2k_{nF}$  which is incommensurate with $2k_F$.
In addition, at $T_o$ the derivative of the correlation length is discontinuous.
From Fig. \ref{lead} we can infer that $T_o(0)=\infty$ and
$T_o(\infty)=0$  with the best fit given by $T_o(\gamma)\sim\gamma^{-0.55}$.
\begin{figure}[h]
\includegraphics[width=\linewidth]{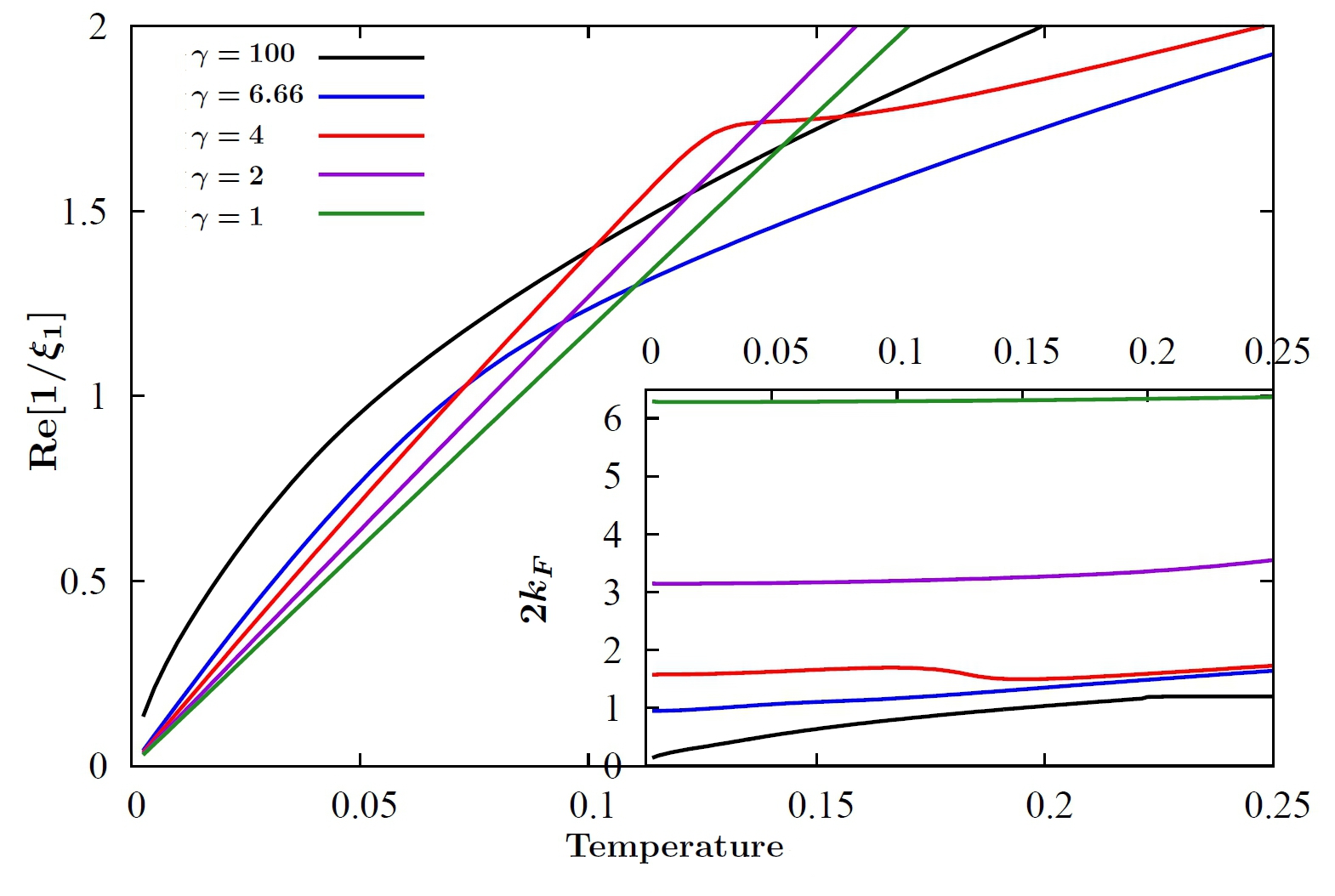}
\caption{(Color online)
$\mbox{Re } [1/\xi_{1}]$ with $\xi_{1}$ the next-leading correlation length as a function of temperature for
various values of $\gamma$.
Inset: $2k_F=\mbox{Im } [1/\xi_{1}]$  for the same parameters. (All quantities in units of $1/l_0$ and $T_0$.)}
\label{next}
\end{figure}
Regarding this last statement it is instructive to look at the asymptotic
expansion (Eq.~(\ref{i3})) of $\langle \rho(x)\rho(0)\rangle_T$ in the
$\gamma=\infty $ limit from which it is easy to see that in this case the
leading term of the correlator is oscillatory at all non-zero temperatures.
 We would like to emphasize the peculiar nature of
  this oscillatory crossover which contrasts with the more common case in
  which the next leading correlation length becomes dominant (see
  below). In the case at hand $\xi_0$ still remains
  dominant but acquires an imaginary part which gives incommensurate
  oscillations.

The next-leading correlation lengths are obtained considering $k_1^+$ in
the second quadrant and $k_1^-$ in the fourth quadrant (denoted by $\xi_{1}$)
or $k_1^+$ in the first quadrant and $k_1^-$ in the third quadrant
(denoted by $\xi_{-1}$). At low-T these correlation lengths reproduce the asymptotic
behavior of the $l=\pm 1$ terms in the TLL/CFT expansion (\ref{inte1}). In
Fig. \ref{next} we show graphs of $1/\xi_{1}$ for some relevant values of
$\gamma$. We notice that the region of validity for the TLL/CFT predictions
(linear dependence on temperature of the correlation lengths and constancy
of the wavevector) decreases with $\gamma$. Also we can clearly see that for
$\gamma=4,$ $\mbox{Re} [1/\xi_{1}]$ develops a ``shoulder" in the region of
temperatures for which the conformal predictions break down. However, it should
be stressed that the derivative  remains continuous  unlike the case of the
leading correlation length.

\begin{figure}[h]
\includegraphics[width=\linewidth]{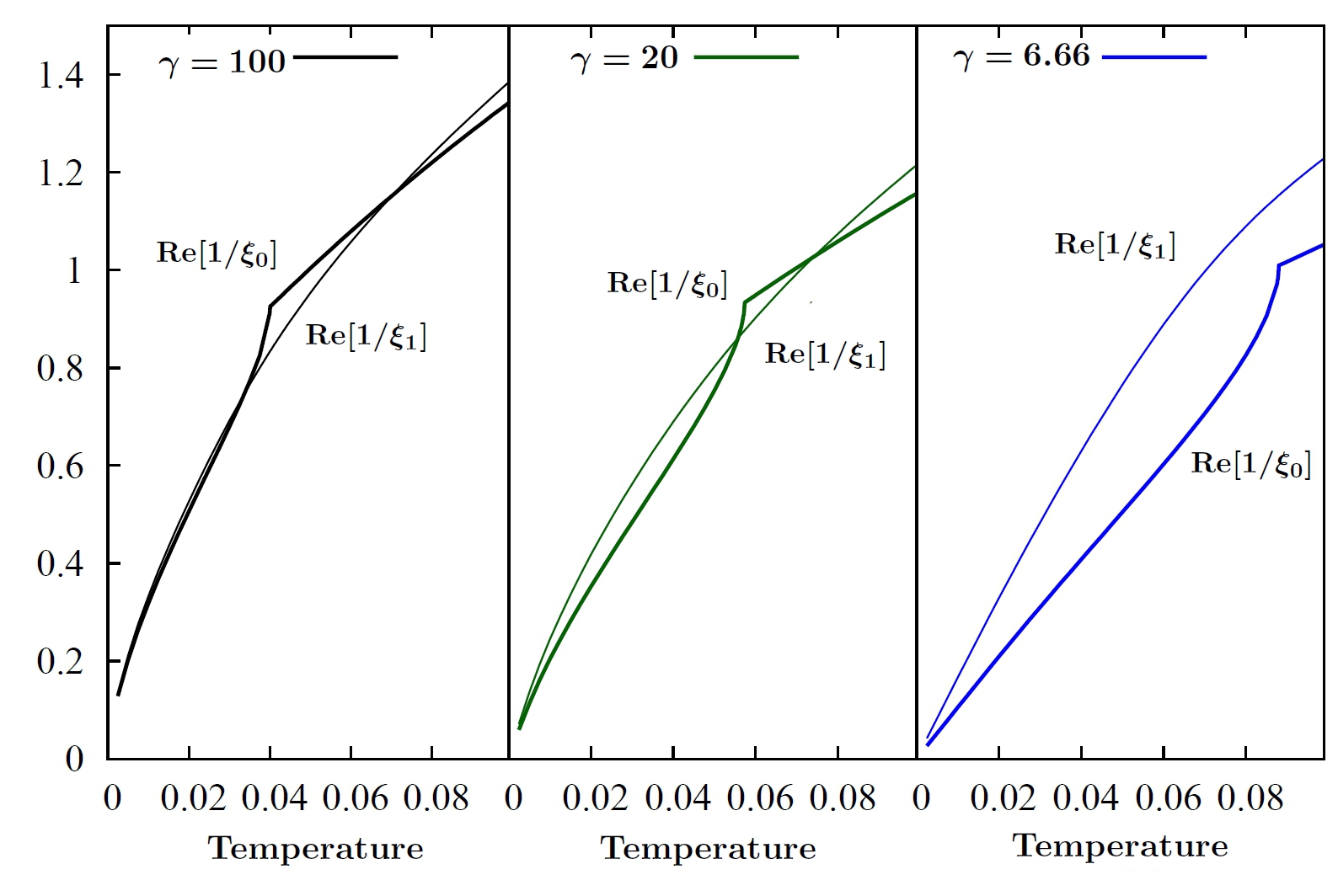}
\caption{(Color online)
$\mbox{Re } [1/\xi_0]$ (thick line) and $\mbox{Re } [1/\xi_{1}]$ (thin line)  as functions of temperature for
various values of $\gamma$. For $\gamma=20$ and $\gamma=100$ a complex crossover scenario can be seen.
(All quantities in units of $1/l_0$ and $T_0$.)
}
\label{both}
\end{figure}

\begin{figure}[h]
\includegraphics[width=\linewidth]{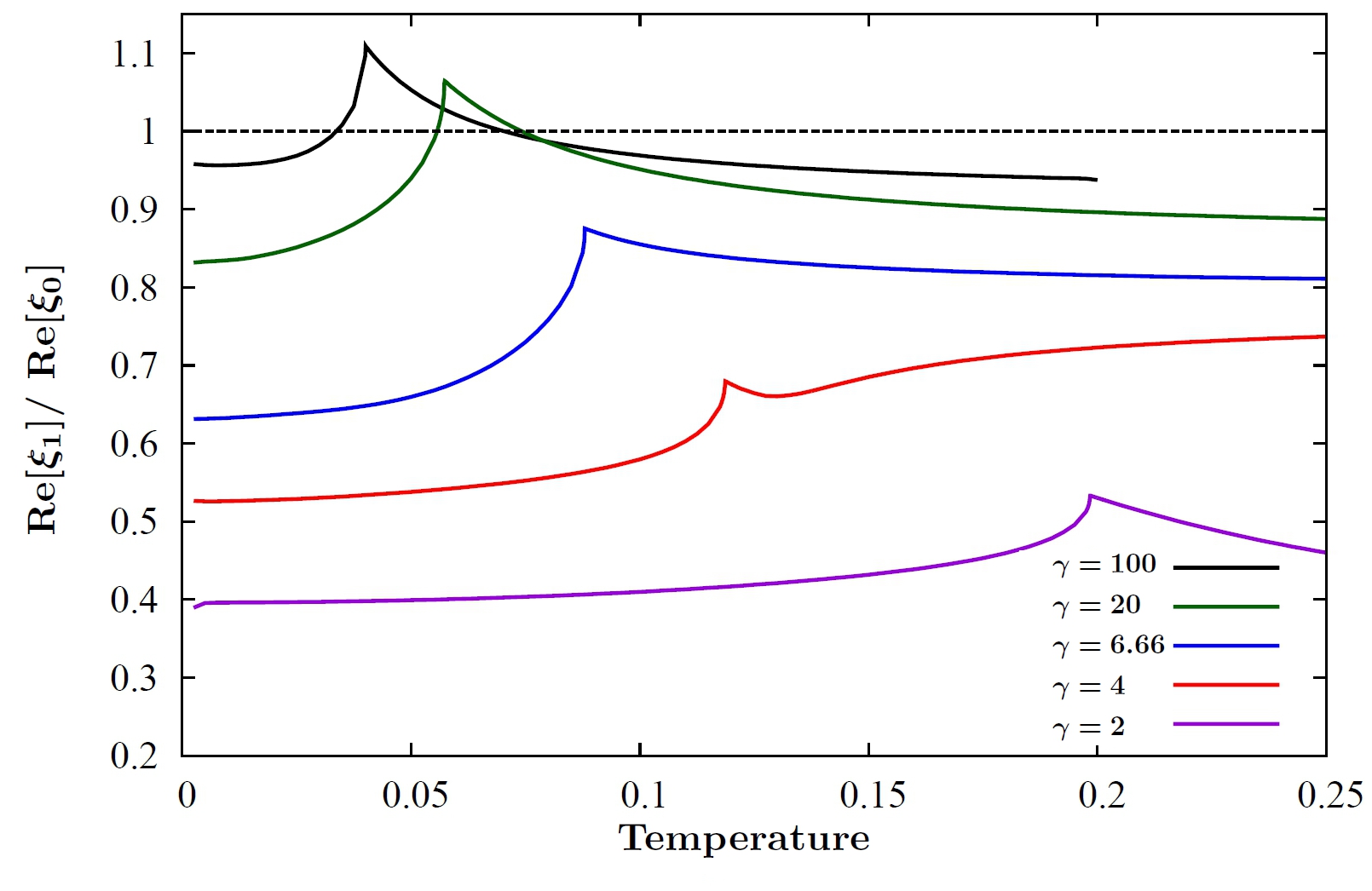}
\caption{(Color online)
$\mbox{Re } [\xi_{1}]/\mbox{Re } [\xi_0]$   as a function of temperature for
various values of $\gamma$. For $\gamma=20$ and $\gamma=100$ in the region of temperatures
where this ratio is greater than one $\xi_{1}$ becomes the leading correlation length.
(Temperature in units of  $T_0$.)
}
\label{ratio}
\end{figure}

Plotting $1/\xi_0$ and $1/\xi_{1}$ in the same graph as in Fig. \ref{both} (see also Fig. \ref{ratio})
reveals an extremely interesting phenomenon. For weak interactions  we have
$\mbox{Re}[1/\xi_0]<\mbox{Re}[1/\xi_{1}]$ for all temperatures but for stronger
interactions a complex crossover scenario emerges. We find that for large $\gamma$
we can distinguish three distinct intervals of temperature for which the asymptotic
behavior of $\langle \rho(x)\rho(0)\rangle_T$ is different. For $T\in(0,T_{lc}(\gamma))$ where $T_{lc}(\gamma)$
is  the lower crossover temperature the leading correlation length is $\xi_0$ which
is real for all temperatures in this interval. $\xi_{1}$ becomes dominant for
$T\in(T_{lc}(\gamma),T_{hc}(\gamma))$ interval in which also lies $T_o$ the temperature
for  which $\xi_0$ acquires a nonzero imaginary part. Here $\langle \rho(x)\rho(0)\rangle_T$ is oscillatory
and exponentially decreasing with leading correlation length $\xi_{1}$. Finally,
for $T>T_{hc}(\gamma)$ we have another crossover with $\xi_0$ (for which
$\mbox{Im}[1/\xi_0]\ne 0$)  characterizing the leading term of the expansion.
It should be emphasized that the description of this crossover scenario is out of
the reach of the TLL/CFT approach or other approximation methods.

{\it Conclusions - } We have performed an extensive numerical investigation of the
large distance asymptotic behavior of the second order correlation function in
the Lieb-Liniger model discovering an extremely complex crossover phenomenon. Our
findings can be summarized as follows. For all strengths of interactions the leading
correlation length develops a nonzero imaginary part for temperatures larger than
a critical temperature $T_o(\gamma)$ with $T_o(0)=\infty$ and
$T_o(\infty)=0$. As we approach the Tonks-Girardeau limit we find
two additional crossovers  in which  $\xi_0$ and $\xi_1$
successively change places as the dominant one.
Interestingly, the crossover phenomena happen when leaving the low-temperature
 dense phase either by increasing the temperature or by decreasing
the density. The incommensurate oscillations stay deep into the gaseous phase
which appears more complex than a true ideal gas. In the gaseous phase Eq.~(\ref{unls})
simplifies as the contribution by the integral vanishes, but the function
$u_i$ is still more complicated than in the ideal gas case.

The crossover to oscillatory behavior of the leading correlation length is
reminiscent of the one observed in the XXZ spin chain \cite{FKM} in the
ferromagnetic region ($-1<\Delta<0$) and zero magnetization. This may somewhat
be expected if we take into account that the Bose gas is a continuum limit of
the XXZ spin chain at $\Delta=-1$, but this holds close to the fully polarized
state. Naturally, the full temperature scenarios are different for the two
systems. However, for cold gases the prospects of finding experimental
signatures of the crossovers are much better than for crystals as a result of
the continuous improvement in the creation and manipulation of such systems.

{\it Acknowledgments - } Financial support from the  VolkswagenStiftung and
the PNII-RU-TE-2012-3-0196 grant  of the Romanian National Authority for
Scientific Research is gratefully acknowledged.

\appendix

\section{Asymptotic behavior in the Tonks-Girardeau limit}\label{aTG}

Deriving the large-distance asymptotic behavior of the density correlator
in the Tonks-Girardeau limit
\be\label{i22}
\langle:\rho(x)\rho(0):\rangle_T=n^2-\frac{1}{4\pi^2}\left(\inti e^{i kx}\vartheta(k)
\, dk\right)^2\, ,\ \ \
\ee
reduces to the asymptotic investigation of the Fourier transform of the Fermi distribution
\be\label{s1}
\inti \frac{e^{ik x}}{1+e^{(k^2-\mu)/T}}\, dk\, .
\ee
We introduce the notation  $f(k|x)=e^{i k x}/(1+e^{(k^2-\mu)/T}).$ $f(k|x)$ is a meromorphic
function on the complex plane with poles  given by the solutions of the  equation
$k^2-\mu=i\pi (2s+1) T\, ,\  s=0,\pm 1,\cdots$. The solutions of this equation are
\begin{align}\label{kk}
k_r^\pm(s)&=[(\alpha(s)+\mu)^{1/2}\pm i \left(\alpha(s)-\mu\right)^{1/2}]/\sqrt{2}\, ,\nonumber\\
k_l^\pm(s)&=[-(\alpha(s)+\mu)^{1/2}\pm i \left(\alpha(s)-\mu\right)^{1/2}]/\sqrt{2}\, ,
\end{align}
with $\alpha(s)=\sqrt{\mu^2+(2s+1)^2\pi^2T^2}$. For $x>0$ (it is sufficient to consider
only this case because $\langle \rho(x)\rho(0)\rangle_T=\langle \rho(0)\rho(x)\rangle_T$),
$f(k|x)$ vanishes at infinity in the upper  half-plane which means that
\[
\inti f(k|x) dk=\int_{\mathcal{C}} f(k|x) dk\, ,
\]
with $\mathcal{C}$ a complex contour which contains the real axis and a semicircle extending
at infinity in the upper half plane. Using this identity and Cauchy's residue theorem we find
\[
\inti f(k|x) dk= 2\pi i\sum_{s=0}^\infty \mbox{Res}[ f(k_l^+(s)|x)+f(k_r^+(s)|x)]\, ,
\]
where we have used the fact that the poles in the upper half-plane of the
function $f(k|x)$ are $k_l^+(s)$ and $k_r^+(s)$ with $s=0,1,\cdots$  and
residues $\mbox{Res} f(k_{l,r}^+(s)|x)=-T e^{i k_{l,r}^+ x}/(2 k_{r,l}^+)$. Taking only the
$s=0$ terms and squaring we obtain Eq.~(\ref{i3}).


\begin{thebibliography}{99}
\bibitem{units}  In order to restore physical units we introduce a length scale $l_0$ via
                 $c=m g l_0/\hbar^2$ where $m$ is the mass of  particles and $g$ the contact
                 interaction strength. Then, the units of temperature, reciprocal correlation
                 length and wavenumber  are $T_0=\hbar^2/(2 m l_0^2 k_B)$ and $1/l_0$.
\end{thebibliography}
\end{document}